\documentclass[12pt]{article}
\usepackage{graphicx}
\catcode`\@=11
\def\numberbysection{\@addtoreset{equation}{section}
\def\theequation{\thesection.\arabic{equation}}}

\numberbysection

\newcommand{\be}{\[}
\newcommand{\bea}{\begin{eqnarray*}}
\newcommand{\beq}{\begin{equation}}
\newcommand{\beqa}{\begin{eqnarray}}
\newcommand{\ee}{\]}
\newcommand{\eea}{\end{eqnarray*}}
\newcommand{\eeq}{\end{equation}}
\newcommand{\eeqa}{\end{eqnarray}}
\newcommand{\abs}[1]{\vert#1\vert}
\newcommand{\bigmean}[1]{\left\langle#1\right\rangle}
\renewcommand{\d}{{\rm d}}
\newcommand{\dpar}{{\partial}}
\newcommand{\e}{{\rm e}}
\newcommand{\eps}{\varepsilon}
\newcommand{\eq}{{\rm eq}}
\newcommand{\frad}[2]{\displaystyle{\displaystyle#1\over\displaystyle#2}}
\newcommand{\lam}{\lambda}
\newcommand{\mean}[1]{\langle#1\rangle}
\newcommand{\s}{\sigma}
\newcommand{\sep}{{\rm SEP}}
\newcommand{\tot}{\leftrightarrow}
\newcommand{\w}{\widehat}
\newcommand{\D}{\Delta}
\newcommand{\K}{{\cal K}}
\renewcommand{\H}{{\cal H}}
\newcommand{\Lam}{\Lambda}
\renewcommand{\P}{{\cal P}}
\renewcommand{\S}{{\cal S}}
\newcommand{\T}{{\cal T}}
\newcommand{\1}{{\bf 1}}

\begin{document}
\centerline{\Large\bf Anomalous self-diffusion in the ferromagnetic}
\vspace{.3cm}
\centerline{\Large\bf Ising chain with Kawasaki dynamics}
\vspace{1.6cm}
\centerline{\large
C.~Godr\`eche$^{a,b,}$\footnote{godreche@spec.saclay.cea.fr}
and J.M.~Luck$^{c,}$\footnote{luck@spht.saclay.cea.fr}}
\vspace{1cm}
\noindent $^a$Dipartimento di Fisica, Universit\`a di Roma ``La Sapienza''
and Center for Statistical Mechanics and Complexity, INFM Roma 1,
Piazzale Aldo Moro 2, 00185 Roma, Italy

\noindent $^b$Service de Physique de l'\'Etat Condens\'e,
CEA Saclay, 91191 Gif-sur-Yvette cedex, France

\noindent $^c$Service de Physique Th\'eorique\footnote{URA 2306 of CNRS},
CEA Saclay, 91191 Gif-sur-Yvette cedex, France
\vspace{1cm}

\begin{abstract}
We investigate the motion of a tagged spin in a ferromagnetic Ising chain
evolving under Kawasaki dynamics.
At equilibrium, the displacement is Gaussian,
with a variance increasing as $A\,t^{1/2}$.
The temperature dependence of the prefactor $A$ is derived exactly.
At low temperature, where the static correlation length $\xi$ is large,
the mean square displacement grows as
$(t/\xi^2)^{2/3}$ in the coarsening regime,
i.e., as a finite fraction of the mean square domain length.
The case of totally asymmetric dynamics,
where $(+)$ (resp.~$(-)$) spins move only to the right (resp.~to the left),
is also considered.
In the steady state, the displacement variance grows as $B\,t^{2/3}$.
The temperature dependence of the prefactor~$B$ is derived exactly,
using the Kardar-Parisi-Zhang theory.
At low temperature, the displacement variance grows as
$t/\xi^2$ in the coarsening regime,
again proportionally to the mean square domain length.
\end{abstract}

\vfill
\noindent P.A.C.S.: 05.70.Ln, 64.60.My, 75.40.Gb.

\newpage
\section{Introduction}

The purpose of this work is to investigate
the random displacement of a tagged spin in an Ising chain
evolving under Kawasaki dynamics~\cite{kawa} from a disordered,
unmagnetized initial configuration.
Many features of the one-dimensional kinetic Ising-Kawasaki model
are by now
well understood~\cite{cs,cks,cb,skr,bray},
although its complete analytical description is still lacking.
In Kawasaki dynamics, only pairs of opposite spins are flipped,
so that the magnetization is locally conserved.
As a consequence, the motion of an individual spin
can be traced throughout the history of the system.

We aim at characterizing the asymptotic
subdiffusive nature of this displacement,
both at equilibrium and in the low-temperature coarsening regime,
first for symmetric, then for asymmetric dynamics.
We use both analytical reasoning and numerical simulations.
In particular,
identifying~$(+)$ (resp.~$(-)$) spins with particles (resp.~holes),
the Ising-Kawasaki chain is recognized as a one-dimensional stochastic
lattice gas, where particles experience hard-core repulsion,
as well as short-ranged attractive interactions
originating from the Hamiltonian~(\ref{ham}) of the spin system.
This identification allows us to borrow concepts and methods from the realm
of interacting particle systems~\cite{ips1,ips2,ips3,ips4,ips5,ips6}.
Finally, this study gives some hindsight into the phenomenon of cage effect,
a subject of much current interest in glassy systems~\cite{hetero}.

Consider a ferromagnetic chain of Ising spins $\s_n=\pm1$ ($n=1,\dots,N$),
with Hamiltonian
\beq
\H=-\sum_n\s_n\s_{n+1},
\label{ham}
\eeq
and periodic boundary conditions.
Since individual spins can be traced throughout the evolution,
a label $k=1,\dots,N$ can be attributed to each spin.
The algebraic displacement $r_k(t)$ of spin number $k$ at time $t$
has an obvious definition: it is initialized at $r_k(0)=0$,
and then increased (resp.~decreased) by one unit
whenever spin number~$k$ moves one step to the right~(resp.~to the left).
This definition of displacement wraps around the sample,
thanks to the periodic boundary conditions.
We shall be mostly interested in moments of the displacement,
defined as averages over all the thermal histories,
for a given tagged spin~$k$ in a given initial configuration.
This is indeed the right procedure in the asymmetric case.
We shall successively consider symmetric~(Section~3)
and asymmetric~(Section~4) Kawasaki dynamics,
both at equilibrium and out of equilibrium.
Section~5 contains a discussion, and a summary of our main results,
listed in Table~\ref{res}.

\section{Ising-Kawasaki chain}

Let us summarize the main features of the model which will be needed hereafter.

\subsection{Symmetric dynamics}

We choose the heat-bath rule to define the stochastic dynamics,
because of its peculiar property that the steady-state is independent
of the applied electric field in the asymmetric case~\cite{kls,ips3}
(see below).
For a pair of opposite spins ($\s_n+\s_{n+1}=0$),
the move $(\s_n\to-\s_n,\s_{n+1}\to-\s_{n+1})$ is realized with probability
\beq
W(\delta\H)=\frac{1}{\exp(\beta\delta\H)+1},
\label{heat}
\eeq
where $\delta\H=2(\s_{n-1}\s_n+\s_{n+1}\s_{n+2})=0,\pm4$
is the energy difference between the configurations after and before the move.
The heat-bath rule~(\ref{heat}) obeys detailed balance
with respect to the Hamiltonian~(\ref{ham}) at temperature $T=1/\beta$,
i.e., $W(\delta\H)=W(-\delta\H)\exp(-\beta\delta\H)$.
The moves corresponding to the three possible values
of the difference $\delta\H$ are listed in Table~\ref{movsym},
with the corresponding acceptance probabilities.

\begin{table}[ht]
\begin{center}
\begin{tabular}{|l|c|c|c|}
\hline
type&$\delta\H$&acc.~probability&moves\\
\hline
condensation&$-4$&$\e^{4\beta}/(\e^{4\beta}+1)$
&$\matrix{-+-\,+\to--+\,+\cr+-+\,-\to++-\,-}$\\
\hline
diffusion&0&$1/2$&$\matrix{++-\,+\tot+-+\,+\cr--+\,-\tot-+-\,-}$\\
\hline
evaporation&$+4$&$1/(\e^{4\beta}+1)$
&$\matrix{++-\,-\to+-+\,-\cr--+\,+\to-+-\,+}$\\
\hline
\end{tabular}
\caption{\small Types of moves in symmetric Kawasaki dynamics,
and corresponding acceptance probabilities with heat-bath rule.}
\label{movsym}
\end{center}
\end{table}

At finite temperature, the three types of moves take place,
and the system relaxes to thermal equilibrium.
The equilibrium state is characterized by independent
bond variables $s_n=\s_n\s_{n+1}=\pm1$, with
\beq
s_n=\left\{\matrix{
-1&\hbox{with probability}&\Pi,\hfill\cr
+1&\hbox{with probability}&1-\Pi,\hfill
}\right.
\label{bonds}
\eeq
where
\be
\Pi=\frac{1}{\e^{2\beta}+1}
\ee
is the density of domain walls.
The mean energy density is therefore $E=-\mean{s_n}=2\Pi-1=-\tanh\beta$,
and spin correlations read
\be
\mean{\s_0\s_n}=(\tanh\beta)^{\abs{n}}=\exp(-\abs{n}/\xi),
\ee
where the equilibrium correlation length,
$\xi=1/\abs{\!\ln\tanh\beta}=1/\abs{\!\ln(1-2\Pi)}$,
diverges as
\be
\xi\approx\frac{\e^{2\beta}}{2}
\ee
at low temperature.
Finally, the reduced susceptibility reads
\beq
\chi=\sum_n\mean{\s_0\s_n}=\e^{2\beta}.
\label{chi}
\eeq

At zero temperature, evaporation moves are not permitted.
The system gets blocked into metastable configurations,
consisting of domains of at least two parallel spins.
Such metastable configurations cannot evolve
under the sole effect of condensation and diffusion moves.
A study of the statistics of the metastable configurations
reached by zero-temperature dynamics can be found in~\cite{uskawa}.
In particular the blocking time is rather short,
increasing as $(\ln N)^3$ for a system of $N$ spins.

At low but non-zero temperature,
evaporation moves take place with the small probability
\beq
\eps=\frac{1}{\e^{4\beta}+1}\approx\1/(4\xi^2)\ll1.
\label{omeg}
\eeq
Starting from a disordered initial configuration,
the system is first attracted into a metastable configuration,
from where it escapes on the slow time scale $\eps t\sim t/\xi^2$.
For $t/\xi^2\gg1$, the system enters a self-similar coarsening regime,
where the typical domain length grows as
$L\sim(t/\xi^2)^{1/3}$~\cite{cs,cks,bray} [see~(\ref{lpower})].
Finally, the domain length saturates to its equilibrium value
\be
L_\eq=\frac{2}{1+E}=\frac{1}{\Pi}=\e^{2\beta}+1\approx2\xi.
\ee
The corresponding equilibration time scales as
\beq
\tau_\eq\sim\xi^5.
\label{tausym}
\eeq

\subsection{Asymmetric dynamics}

Assume that $(+)$ spins (resp.~$(-)$ spins) are positively
(resp.~negatively) charged.
In the presence of an electric field,
they move preferentially to the right (resp.~to the left).
For the sake of simplicity,
consider the limit of totally asymmetric dynamics~\cite{cb,skr},
where only $+\,-\to-\,+$ moves are allowed.
These moves are listed in Table~\ref{movbias},
with the corresponding acceptance probabilities.

\begin{table}[ht]
\begin{center}
\begin{tabular}{|l|c|c|c|}
\hline
type&$\delta\H$&acc.~probability&move\\
\hline
condensation&$-4$&$\e^{4\beta}/(\e^{4\beta}+1)$&$\matrix{-+-\,+\to--+\,+}$\\
\hline
conduction&0&$1/2$&$\matrix{++-\,+\to+-+\,+\cr-+-\,-\to--+\,-}$\\
\hline
evaporation&$+4$&$1/(\e^{4\beta}+1)$&$\matrix{++-\,-\to+-+\,-}$\\
\hline
\end{tabular}
\caption{\small Types of moves in totally asymmetric Kawasaki dynamics,
and corresponding acceptance probabilities with heat-bath rule.}
\label{movbias}
\end{center}
\end{table}

Under asymmetric dynamics at finite temperature,
the driven system reaches a non-equilibrium steady state,
where $(+)$ spins (resp.~$(-)$ spins) are advected
with a constant drift velocity $V$ (resp.~$-V$),
to be derived below~[see~(\ref{kpzv})].
All the static quantities, such as equal-time correlations,
coincide with their equilibrium values~\cite{kls,ips3}
for a whole class of stochastic dynamical rules,
including the heat-bath rule~(\ref{heat}).
This does not hold with the Metropolis rule.

At zero temperature,
the system again gets blocked into metastable configurations,
consisting of domains of at least two parallel spins.
The statistics of metastable configurations
is different from that obtained in the symmetric case.
This difference is explicitly apparent
in the special case of restricted zero-temperature dynamics
(descent dynamics), where only condensation moves are allowed.
In this situation, analytical results can be obtained,
both for the symmetric case~\cite{plk,usrsa} and the asymmetric one~\cite{cp}.

At low temperature,
the system exhibits self-similar domain growth for $t/\xi^2\gg1$,
with a typical domain length increasing as $L\sim(t/\xi^2)^{1/2}$~\cite{cb,skr}
[see~(\ref{blpower})].
The equilibration time scales as
\beq
\tau_\eq\sim\xi^4.
\label{tauasym}
\eeq

\section{Symmetric dynamics}

\subsection{Mean square displacement at equilibrium}

Assume that the system is at thermal equilibrium,
evolving under symmetric Kawasaki dynamics.

Consider first the case of infinite temperature.
Then all the allowed moves have equal acceptance probabilities $1/2$.
Particles (i.e., $(+)$ spins) hop along the chain,
with the hard-core constraint that sites are occupied by at most one particle.
The motion of holes (i.e., $(-)$ spins) follows by complementarity.
The process thus defined is known as
the symmetric exclusion process (SEP)~\cite{ips1,ips4,ips5,ips6}.
The particle density~$\rho$ is fixed by the magnetization of the initial
configuration: $M=\mean{\s_n}=1-2\rho=0$, hence $\rho=1/2$.
The displacement of individual particles is asymptotically Gaussian,
with a variance increasing as $t^{1/2}$~\cite{sep,ips1}.

We now show that the same holds for the system at equilibrium
at any finite temperature,
i.e., we predict a subdiffusive displacement of tagged particles,
with a symmetric Gaussian dispersion profile, whose variance scales as
\beq
\mean{r^2}\approx A\,t^{1/2}.
\label{lawse}
\eeq
The temperature dependence of the amplitude $A$ is given in~(\ref{aew}).

In order to do so,
we use the Edwards-Wilkinson (EW) continuum theory~\cite{ew}.
The starting point of this approach consists
in mapping the problem onto an interface model~\cite{ek1,ek2,ek3},
by defining the height variable
\beq
h_n=\sum_{m=-N}^n\s_m.
\label{hdef}
\eeq
The reference point $n=-N$ does not actually matter,
since only height differences will be considered hereafter.
One has e.g.
\beq
\mean{(h_m-h_n)^2}\approx\abs{m-n}\chi
\label{compress}
\eeq
for large separations $(\abs{m-n}\gg\xi)$ at equilibrium,
where the compressibility~$\chi$ is just the reduced susceptibility
evaluated in~(\ref{chi}).
Height and displacement are related by the following
approximate relationship~\cite{ek1,ek2,ek3}
\beq
h_n(t+\tau)-h_n(t)=-2 N_n(t+\tau,t)\approx-(r(t+\tau)-r(t))_n.
\label{hr}
\eeq
In this expression,
$N_n(t+\tau,t)$ is the net number of particles ($(+)$ spins)
which have crossed the bond $(n,n+1)$ between times $t$ and $t+\tau$,
particles moving to the right (resp.~to the left)
being counted positively (resp.~negatively),
while $(r(t+\tau)-r(t))_n$ is the displacement between times $t$ and $t+\tau$
of particles in the neighborhood of site $n$.
Particles indeed move together almost rigidly,
as the system has a finite static compressibility
$\chi$~[see~(\ref{compress})].

In the continuum limit, the height $h(x,t)\approx-r(x,t)$
obeys the EW equation~\cite{ek1,ek2}
\beq
\frac{\dpar h}{\dpar t}=D\frac{\dpar^2 h}{\dpar x^2}+\eta(x,t),
\label{eweq}
\eeq
where $\eta(x,t)$ is a Gaussian white noise, such that
$\mean{\eta(x,t)\eta(x',t')}=2\D\,\delta(x-x')\delta(t-t')$.
In Fourier space,~(\ref{eweq}) yields
\be
\w h(q,t)=\w h(q,0)\,\e^{-Dq^2t}+\int_0^t\d s\,\w\eta(q,s)\,\e^{-Dq^2(t-s)}.
\ee
We thus obtain the following expression
\be
\mean{(h(x,t+\tau)-h(x,t))^2}
\to 2\D\int_{-\infty}^\infty\frac{\d q}{2\pi}\,\frac{1-\e^{-Dq^2\tau}}{Dq^2}
=2\D\left(\frac{\tau}{\pi D}\right)^{1/2}
\ee
for the height (or displacement) correlation function
at equilibrium $(t\to\infty)$.
The noise intensity $\D$ can be evaluated from the height correlation
at equilibrium,
\be
\mean{(h(x,t)-h(0,t))^2}
\to 2\D\int_{-\infty}^\infty\frac{\d q}{2\pi}\,\frac{1-\cos qx}{Dq^2}
=\frac{\D\,x}{D}.
\ee
Comparing with~(\ref{compress}) yields the identification $\D=D\chi$,
and finally
\beq
A=2\chi\left(\frac{D}{\pi}\right)^{1/2}.
\label{aresew}
\eeq
We now estimate the magnetization diffusion coefficient $D$.
A hierarchy of non-linear kinetic equations
for spin correlation functions can be written
by summing the contributions of the moves listed in Table~\ref{movsym}.
For the magnetization, we obtain
\beq
\frac{\d\mean{\s_n}}{\d t}=S_{n-1}+S_{n+1}-2S_n,
\label{sdif}
\eeq
with
\beq
S_n=\frac{1}{2}\,\mean{\s_n}
+\frac{\e^{4\beta}-1}{4(\e^{4\beta}+1)}
\,\mean{\s_{n-1}\s_n\s_{n+1}-\s_{n-1}-\s_n-\s_{n+1}}.
\label{grands}
\eeq
The nonlinear equation~(\ref{sdif}) cannot be solved in full generality,
except at infinite temperature~\cite{ll}, where it becomes linear.
However, since~(\ref{sdif}) already involves a second-order difference,
it is sufficient for the present purpose
to evaluate the uniform value $S$ of~$S_n$
in the finite-temperature equilibrium state of the Ising chain
with a small uniform magnetization $M=\mean{\s_n}$.
Indeed, (\ref{sdif}) implies
\beq
D=\lim_{M\to0}\frac{S}{M}.
\label{dsm}
\eeq
The explicit calculation of $S$ can be performed by means of
the transfer-matrix formalism.
Introducing a uniform reduced magnetic field $H$, the transfer matrix reads
\be
\T=\pmatrix{\e^{\beta+H}&\e^{-\beta-H}\cr\e^{-\beta+H}&\e^{\beta-H}}.
\ee
The leading eigenvalue of $\T$ and the corresponding eigenvectors read
\be
\Lam=\e^{\beta}\cosh H+\e^{-\beta}(1+\e^{4\beta}\sinh^2H)^{1/2},
\ee
\be
\langle L\vert=\left(\Lam-\e^{\beta-H},\;\e^{-\beta-H}\right),\quad
\vert R\rangle=\pmatrix{\e^{-\beta-H}\cr\Lam-\e^{\beta+H}}.
\ee
We have in particular
\be
M=\frac{\langle L\vert\S\vert R\rangle}{\langle L\vert R\rangle},
\ee
where the spin operator reads
\be
\S=\pmatrix{1&0\cr0&-1},
\ee
hence
\beq
M=\frac{\e^{2\beta}\sinh H}{(1+\e^{4\beta}\sinh^2H)^{1/2}},\qquad
\e^H=\frac{M+\left(M^2+\e^{4\beta}(1-M^2)\right)^{1/2}}
{\e^{2\beta}(1-M^2)^{1/2}}.
\label{mhhm}
\eeq
The equilibrium value of any spin correlation can be expressed
in terms of matrix elements involving the eigenvectors $\langle L\vert$
and $\vert R\rangle$, and the spin operator $\S$.
Using the second expression of~(\ref{mhhm}),
results can be written in terms of $M$ alone.
We have thus
\bea
&&{\hskip -30pt}\mean{\s_{n-1}\s_n\s_{n+1}}
=\frac{\langle L\vert\S\T\S\T\S\vert R\rangle}{\Lam^2\langle L\vert R\rangle}
\cr\cr
&&{\hskip 38pt}
=\frac{\Bigl[8\left(M^2+\e^{4\beta}(1-M^2)\right)^{1/2}
+\e^{4\beta}(\e^{4\beta}-6)(1-M^2)-3-5M^2\Bigr]M}
{(\e^{4\beta}-1)^2(1-M^2)}\cr\cr
&&{\hskip 38pt}
\approx\frac{(\e^{2\beta}-1)(\e^{2\beta}+3)}{(\e^{2\beta}+1)^2}\,M
\quad(M\to0).
\eea
Inserting the latter estimate into~(\ref{grands}), (\ref{dsm}) results in
\be
D=\frac{2}{(\e^{2\beta}+1)(\e^{4\beta}+1)}=2\eps\Pi.
\ee
Finally, using~(\ref{aresew}), we obtain the temperature dependence of
the prefactor $A$:
\beq
A
=\chi\left(\frac{8\eps\Pi}{\pi}\right)^{1/2}
=\left(\frac{8\,\e^{4\beta}}{\pi(\e^{2\beta}+1)(\e^{4\beta}+1)}\right)^{1/2}.
\label{aew}
\eeq

In the infinite-temperature case, we thus recover
$A_\sep=(2/\pi)^{1/2}$~\cite{sep,ips1}.

In the low-temperature regime, the amplitude $A$ vanishes as
\beq
A\approx(8/\pi)^{1/2}\,\e^{-\beta}\approx2(\pi\xi)^{-1/2},
\label{ase}
\eeq
hence the low-temperature scaling form $\mean{r^2}\sim(t/\xi)^{1/2}$.
The latter result is compatible with a smooth crossover,
for times $t\sim\tau_\eq\sim\xi^5$ [see~(\ref{tausym})],
between the behavior~(\ref{rpower}) in the coarsening regime
and the behavior~(\ref{lawse}) in the equilibrium regime.

\begin{figure}[ht]
\begin{center}
\includegraphics[angle=90,width=.6\linewidth]{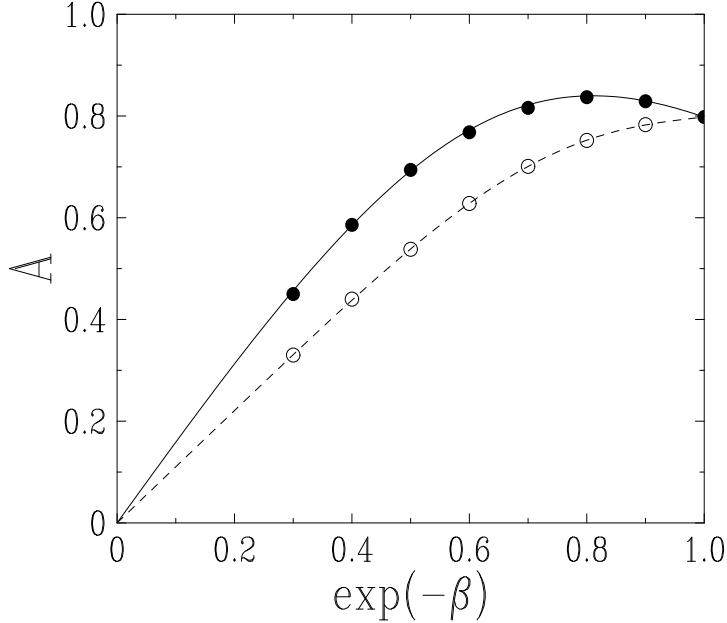}
\caption{\small
Plot of the amplitude $A$ [see~(\ref{lawse})]
of the spin displacement for symmetric equilibrium dynamics,
against $\e^{-\beta}$.
Full symbols: measured amplitude for an equilibrated initial condition
(initial temperature $=$ final one).
Full line: prediction~(\ref{aew}) of EW theory.
Open symbols: measured amplitude for a disordered initial condition
(infinite initial temperature).
Dashed line: polynomial fit to the latter data, meant as a guide to the eye.}
\label{figa}
\end{center}
\end{figure}

In order to test the accuracy of Monte-Carlo numerical simulations,
which will be used more extensively hereafter,
we have checked the prediction~(\ref{aew}) against values of~$A$
obtained by extrapolating numerical results for $\mean{r^2}$
at equilibrium, for various temperatures.
Figure~\ref{figa} shows a plot of $A$ against $\e^{-\beta}$,
both for an unmagnetized equilibrated initial condition
(initial temperature $=$ final one), built recursively using~(\ref{bonds}),
and for an unmagnetized disordered initial condition
(infinite initial temperature).
In the equilibrated case, the measured values of the amplitude
are in perfectly good agreement with~(\ref{aew}).
In the disordered case, $A$ is found to be systematically smaller
than the predicted value~(\ref{aew}).
The amplitude $A$ actually depends continuously on the initial condition,
as can be observed by preparing the system at equilibrium,
at a temperature different from the chosen final temperature.
This dependence on the initial configuration is somewhat surprising,
since the amplitude $A$ characterizes the long-time behavior~(\ref{lawse})
of the displacement of a tagged particle.
Let us mention that data points
for lower temperatures are more and more difficult to obtain,
because the equilibrium regime is reached
for $t\sim\tau_\eq\sim\xi^5\sim\e^{10\beta}$.

\subsection{Mean square displacement in the coarsening regime}

Consider the displacement of a tagged particle
after a quench from an infinite-temperature
disordered initial configuration to a low temperature such that $\xi\gg1$.

The system is first attracted into a metastable configuration,
from where it escapes on the slow time scale $\eps t\sim t/\xi^2$
(see Section~2.1).
Correspondingly $\mean{r^2}$ reaches first a plateau value,
from which it escapes on the same time scale.
For $t/\xi^2\gg1$, the system enters a self-similar coarsening regime,
and $\mean{r^2}$ is observed to grow as $(t/\xi^2)^{2/3}$.
Finally, as the domain length saturates to its equilibrium value,
for $t\sim\tau_\eq\sim\xi^5$,
$\mean{r^2}$ crossovers to its equilibrium behavior described in Section~3.1.

We now provide an explanation for the observed anomalous
diffusion for $t/\xi^2\gg1$.
We use the picture of Cordery et al.~\cite{cs} and Cornell et al.~\cite{cks}
for the low-temperature coarsening regime
(see~\cite{bnk} for further developments).
Consider a $(+)$ spin evaporating from the right boundary of a $(+)$ domain,
of length~$\ell_1$, say.
It then undergoes a free random walk inside the right neighboring
$(-)$ domain, of length $\ell$, until it condensates,
either (i) to the $(+)$ domain from where it had evaporated,
or (ii) to the next~$(+)$ domain to the right, of length $\ell_2$, say.
Event (i) amounts to no domain displacement at all,
while event (ii) amounts to moving
the $(-)$ domain by one lattice spacing to the left,
hence $\ell_1\to\ell_1-1$, and $\ell_2\to\ell_2+1$.
The probability of event~(ii), knowing the spin has evaporated,
is exactly $1/\ell$.
This is a well-known result in the gambler's ruin problem~\cite{feller}.

In other words, integrating over the fast processes
(diffusion and condensation),
one is led to an effective description of the coarsening process
in terms of diffusion and annihilation of domains,
where each domain performs a random walk independently of the others,
with a diffusion constant
\beq
D(\ell)\approx\frac{\eps}{\ell}\approx\frac{1}{4\xi^2\ell}
\label{dofell}
\eeq
for a domain of length $\ell$.
When a domain length shrinks down to zero,
the two neighboring domains coalesce.
Since a domain of length~$L$ evolves on a time scale
of order $t\sim L^2/D(L)\sim\xi^2L^3$,
this picture predicts the power-law growth~\cite{cs,cks,bray}
\beq
L\approx a_L(t/\xi^2)^{1/3},
\label{lpower}
\eeq
where the amplitude $a_L$ is some non-universal number of order unity.
Note that the range of validity
of the coarsening regime ($1\ll L\ll\xi$, or $\xi^2\ll t\ll\tau_\eq\sim\xi^5$)
is larger and larger at lower temperatures.
In this regime, the distribution of domain lengths has the scaling form
\beq
P(\ell)=\frac{1}{L}\,f\!\left(\frac{\ell}{L}\right).
\label{dom}
\eeq

Coming back to the displacement of a tagged particle,
the scenario of domain diffusion and annihilation
described above demonstrates that a tagged particle (i.e., a $(+)$ spin)
exhibits two kinds of motion in the coarsening regime:

\begin{enumerate}

\item[(a)] Most of the time, the tagged particle remains inside a $(+)$ domain.
This domain, and the tagged particle contained in it, perform a random walk,
with an instantaneous diffusion constant given by~(\ref{dofell}).
These moves generate a Gaussian displacement, such that
\be
\frad{\d\mean{r^2}}{\d t}\approx2\mean{D(\ell)}
\approx\frac{1}{2\xi^2}\bigmean{\frac{1}{\ell}}.
\ee
The scaling laws~(\ref{dom}) and~(\ref{lpower})
imply $\mean{1/\ell}\sim1/L\sim(t/\xi^2)^{-1/3}$.
As a consequence, the number of moves of type (a) and the corresponding
mean square displacement up to time $t$
both scale as $N_{({\rm a})}\sim\mean{r^2}\sim L^2\sim(t/\xi^2)^{2/3}$.

\item[(b)] From time to time,
the tagged particle evaporates from the boundary
of the $(+)$ domain to which it belongs, and condensates to a neighboring
domain.
The displacement of the particle in such a move is the length $\ell$
of the $(-)$ domain which is crossed.
It is therefore distributed according to the law~(\ref{dom}).
Furthermore, for a given tagged particle, successive such displacements
alternate in sign,
because the particle can only bounce between two neighboring $(+)$ domains.
The number of such moves scales as $N_{({\rm b})}\sim L\sim(t/\xi^2)^{1/3}$,
and the corresponding random displacement
scales as $\mean{r^2}\sim L^2\sim(t/\xi^2)^{2/3}$.

\end{enumerate}

\begin{figure}[ht]
\begin{center}
\includegraphics[angle=90,width=.6\linewidth]{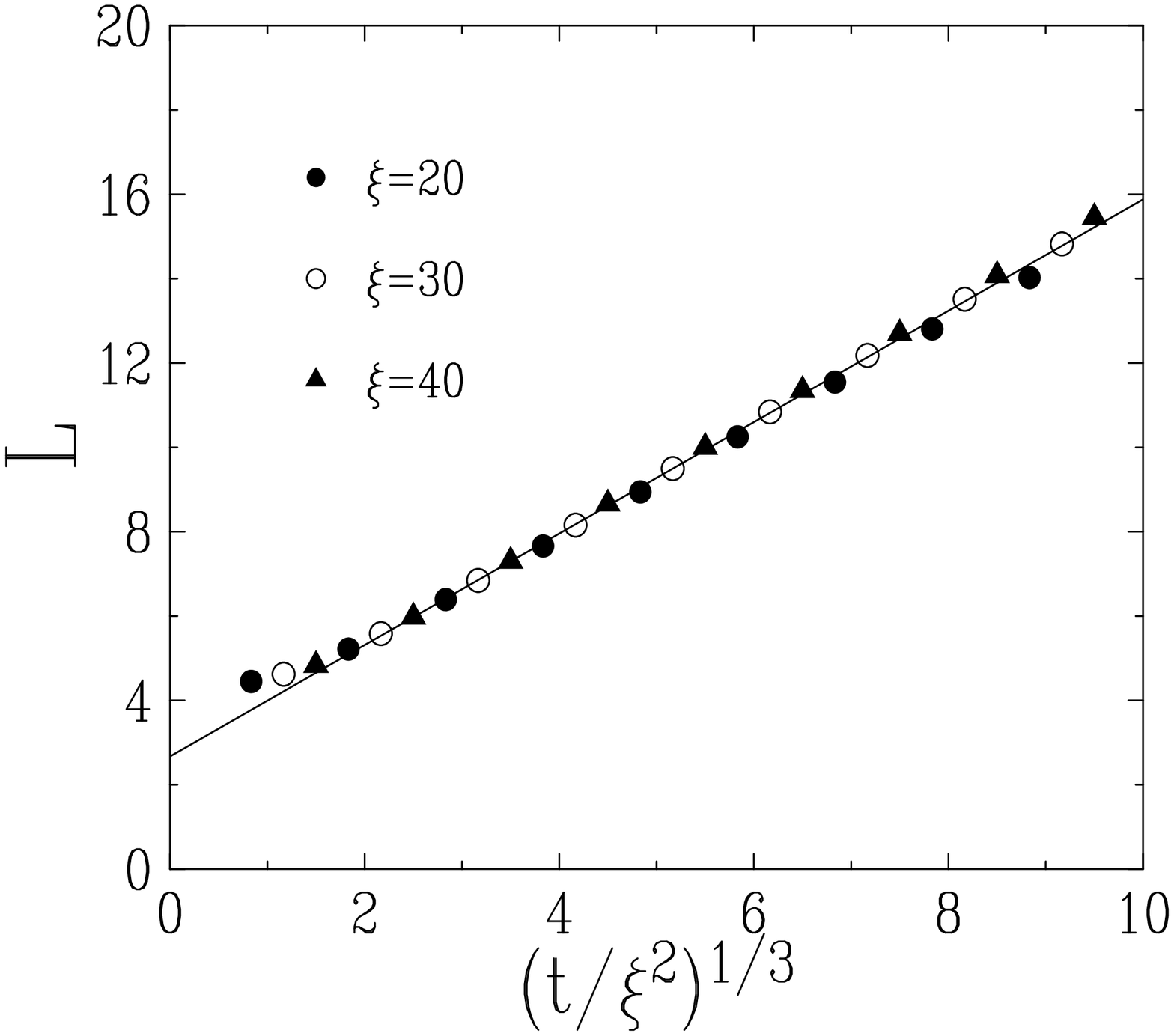}
\caption{\small
Plot of the mean domain length $L$
in the coarsening regime with symmetric dynamics, against $(t/\xi^2)^{1/3}$.
Symbols: data for several values of $\xi$.
Full straight line: least-square fit of the data
(first three points excluded), confirming~(\ref{lpower}).
The slope yields $a_L\approx1.34$.}
\label{figs1}
\end{center}
\end{figure}

To sum up, moves of either type bring similar contributions to $\mean{r^2}$.
We thus predict the power law
\beq
\mean{r^2}\approx a_r^2(t/\xi^2)^{2/3},
\label{rpower}
\eeq
where the amplitude $a_r$ is of order unity.
As a consequence, the dimensionless ratio $\mean{r^2}/L^2$ approaches
a non-trivial universal limit value
\beq
Q=\lim_{t/\xi^2\to\infty}\frac{\mean{r^2}}{L^2}=(a_r/a_L)^2
\label{qq}
\eeq
in the late stages of the coarsening regime.
The sum of moves of type (a) is asymptotically Gaussian.
The distribution of the sum of moves of type (b) is not known a priori,
although it might be expected to be Gaussian as well.

\begin{figure}[ht]
\begin{center}
\includegraphics[angle=90,width=.6\linewidth]{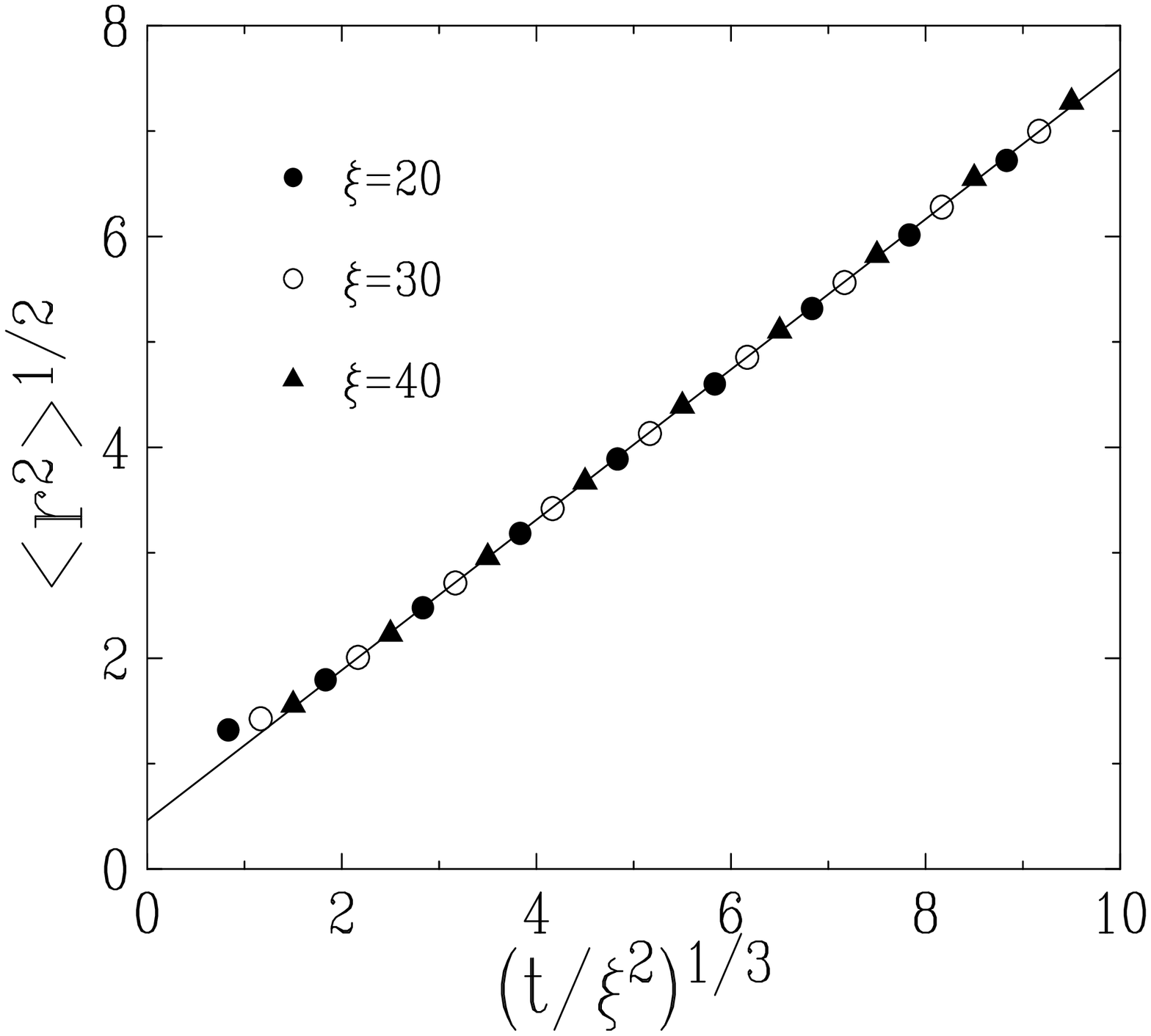}
\caption{\small
Plot of the root-mean-square displacement $\mean{r^2}^{1/2}$
in the coarsening regime with symmetric dynamics, against $(t/\xi^2)^{1/3}$.
Symbols: data for several values of $\xi$.
Full straight line: least-square fit of the data
(first three points excluded), confirming~(\ref{rpower}).
The slope yields $a_r\approx0.71$.}
\label{figs2}
\end{center}
\end{figure}

We have checked the above predictions,
and tested the Gaussian character of the distribution of the displacement,
by means of numerical simulations.
Figures~\ref{figs1} and~\ref{figs2} show plots of the mean domain length $L$,
obtained by measuring the energy density $E=-1+2/L$,
and of the root-mean-square displacement
$\mean{r^2}^{1/2}$, in the coarsening regime,
against the scaling time variable $(t/\xi^2)^{1/3}$,
for several values of $\xi\gg1$.
The data are averaged over 1000 independent
systems of 2000 spins, for times up to $t=1000\,\xi^2$.
A local smoothing procedure has been applied to the raw data.
The observed data collapse and linear behavior confirm the
power laws~(\ref{lpower}), (\ref{rpower}), while least-square fits yield
\be
a_L\approx1.34,\qquad a_r\approx0.71,\qquad Q\approx0.28.
\ee

\begin{figure}[ht]
\begin{center}
\includegraphics[angle=90,width=.6\linewidth]{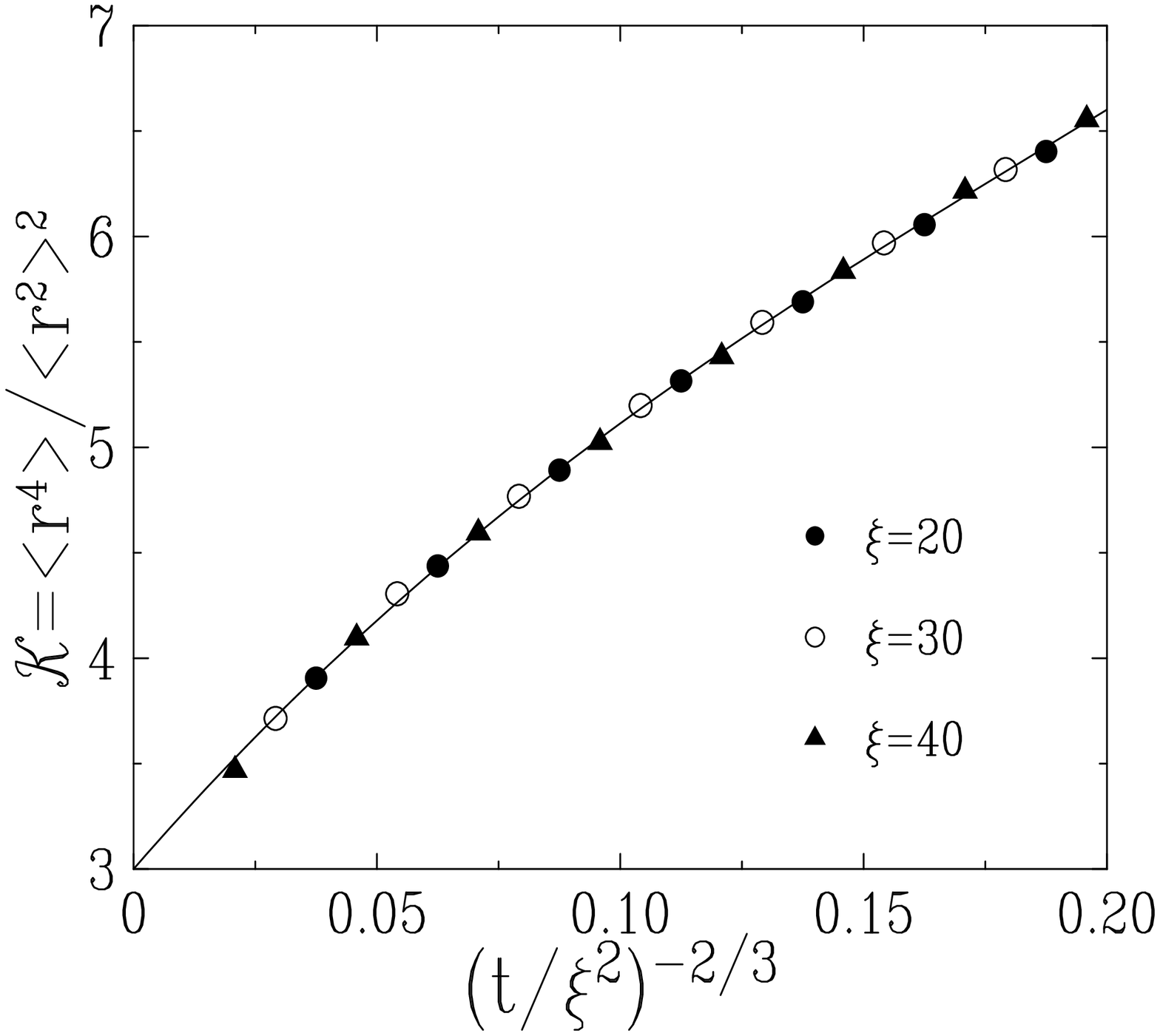}
\caption{\small
Plot of the kurtosis $\K=\mean{r^4}/\mean{r^2}^2$
in the coarsening regime with symmetric dynamics,
against $(t/\xi^2)^{-2/3}$.
Symbols: data for several values of $\xi$.
Full line: polynomial fit demonstrating a smooth convergence
to the limit value $\K=3$ of the Gaussian law.}
\label{figs3}
\end{center}
\end{figure}

Figure~\ref{figs3} shows a plot of the kurtosis
$\K=\mean{r^4}/\mean{r^2}^2$, against $(t/\xi^2)^{-2/3}$.
The data smoothly extrapolate toward the value $\K=3$ of the Gaussian law,
with a correction proportional to $(t/\xi^2)^{-2/3}\sim1/L^2$.
This observation strongly suggests that the asymptotic law of the displacement
is Gaussian.

\section{Asymmetric dynamics}

\subsection{Displacement variance in the steady state}

The system is assumed to be in the steady state reached
under totally asymmetric Kawasaki dynamics at finite temperature.

Consider first infinite-temperature dynamics.
Particles, that is $(+)$ spins, hop to the right, with a hard-core constraint.
This process is known as the totally asymmetric exclusion process
(ASEP)~\cite{ips1,ips4,ips5,ips6}.
The displacement of individual particles is
asymptotically described~\cite{ek1,ek2}
by the nonlinear Kardar-Parisi-Zhang (KPZ) theory~\cite{kpz}.
This theory predicts that,
{\it for a given tagged spin in a given initial configuration},
the displacement at time~$t$ consists of
a drift with uniform velocity, and a thermal random displacement,
increasing typically as~$t^{1/3}$.
This anomalous, subdiffusive width is masked by usual
diffusive fluctuations in $t^{1/2}$ whenever the displacement
is averaged either over different spins
or over different initial configurations.
These peculiar features have been underlined in~\cite{vanb,mb}.

At finite temperature, we expect the same to hold at long times,
i.e., we predict that the motion of a tagged particle consists of
a deterministic drift and a subdiffusive thermal motion, according to
\beq
\mean{r}\approx Vt,\qquad\mean{r^2}_c\approx B\,t^{2/3}.
\label{lawbe}
\eeq

The temperature dependence of the velocity $V$
and of the amplitude~$B$ are now derived, using the KPZ theory.
The most general KPZ equation reads
\beq
\frac{\dpar h}{\dpar t}
=\mu+\nu\frac{\dpar h}{\dpar x}+D\frac{\dpar^2 h}{\dpar x^2}
+\frac{\lam}{2}\left(\frac{\dpar h}{\dpar x}\right)^2+\eta(x,t).
\label{kpzeq}
\eeq
Unlike~(\ref{eweq}), this equation cannot be solved explicitly.
However, all the quantities of interest can be predicted~\cite{ek1,ek2}
from the sole knowledge of the steady-state current~$J(M)$
for a non-zero magnetization $M=1-2\rho$, where $\rho$ is the particle density.
This current can be evaluated by summing the contributions
of the moves listed in Table~\ref{movbias}.
The corresponding steady-state probabilities can again be derived
by means of the transfer-matrix formalism.
Let $\P_\pm=(\1\pm\S)/2$ be the projectors onto the spin states~$\pm$.
The first move involves the probability $p_{-+-+}$
of observing the pattern $-+-\,+$:
\bea
&&{\hskip -17pt}p_{-+-+}
=\frac{\langle L\vert\P_-\T\P_+\T\P_-\T\P_+\vert R\rangle}
{\Lam^3\langle L\vert R\rangle}\cr\cr
&&{\hskip 16pt}=\frac{(3+M^2+\e^{4\beta}(1-M^2))\!
\left(M^2+\e^{4\beta}(1-M^2)\right)^{1/2}
-(1+3M^2)-3\e^{4\beta}(1-M^2)}{2(\e^{4\beta}-1)^3(1-M^2)}.
\eea
Taking into account all the allowed moves, we obtain
\beqa
&&{\hskip -32.5pt}J(M)=\frac{\e^{4\beta}}{\e^{4\beta}+1}p_{-+-+}
+\frac{1}{2}(p_{++-+}+p_{-+--})+\frac{1}{\e^{4\beta}+1}p_{++--}\cr\cr
&&=\frac{2M^2+\e^{4\beta}(1-M^2)-(1+M^2)
\left(M^2+\e^{4\beta}(1-M^2)\right)^{1/2}}
{2(\e^{8\beta}-1)(1-M^2)}.
\label{jofm}
\eeqa
This expression shows that $J(M)=J_0+J_2M^2+\cdots$ is an even function of $M$,
as expected from the spin symmetry of the dynamics,
i.e., the invariance of the dynamical rules
under the simultaneous flipping of all the spins.

As a consequence of~(\ref{hdef}) and~(\ref{hr}),
we are led to identify $\dpar h/\dpar t\approx-2\,\dpar N/\dpar t\approx-2J$
and $\dpar h/\dpar x\approx M$ in the continuum limit.
Therefore, in~(\ref{kpzeq}) we must have
\be
\mu=-2J_0,\qquad\nu=0,\qquad\lam=-4J_2.
\ee
Furthermore, the zero-magnetization velocity $V$
is such that $J_0=\rho V$ with $\rho=1/2$, hence $V=2J_0=-\mu$.
Equation~(\ref{jofm}) yields the explicit results
\beq
V=\frac{\e^{2\beta}}{(\e^{2\beta}+1)(\e^{4\beta}+1)}
=\chi\eps\Pi,\qquad
\lam=\frac{3\e^{2\beta}-1}{\e^{2\beta}(\e^{2\beta}+1)(\e^{4\beta}+1)}.
\label{kpzv}
\eeq

According to the KPZ scaling theory~\cite{ek1,ek2,kmh},
for a given tagged spin in a given initial configuration,
the displacement fluctuation is asymptotically distributed according~to
\be
r-\mean{r}\approx\left(\frac{\chi^2\lam t}{2}\right)^{1/3}X
=\left(\frac{\e^{2\beta}(3\e^{2\beta}-1)}
{2(\e^{2\beta}+1)(\e^{4\beta}+1)}\right)^{1/3}X.
\ee
The reduced variable $X$ has a universal distribution,
related to the Tracy-Widom law
of the largest eigenvalue of a complex Hermitian random matrix.
This distribution,
which attracted much interest in recent years~\cite{ps},
can be expressed in terms of a solution
of a Painlev\'e transcendental equation.

As a consequence, the reduced central moments
\beq
K_n
=\frac{\mean{(r-\mean{r})^n}}{\mean{r^2}_c^{n/2}}
=\frac{\mean{(X-\mean{X})^n}}{\mean{X^2}_c^{n/2}}
\label{redcm}
\eeq
have non-trivial universal values.
The variance $\mean{X^2}_c=0.63805$,
and the reduced moments $K_3=0.2935$, $K_4=0.1652$,
are known numerically to a high accuracy~\cite{ps}.
Finally, the amplitude $B$ entering~(\ref{lawbe}) reads
\beq
B=\left(\frac{\e^{2\beta}(3\e^{2\beta}-1)}
{2(\e^{2\beta}+1)(\e^{4\beta}+1)}\right)^{2/3}
\underbrace{\mean{X^2}_c}_{\displaystyle 0.63805}\!.
\label{kpzb}
\eeq
In the low-temperature regime, we thus obtain
\beq
V\approx\e^{-4\beta}\approx 1/(4\xi^2),
\qquad B\approx\underbrace{(3/2)^{2/3}\mean{X^2}_c}_{\displaystyle
0.83608}\,\e^{-4\beta/3}
\approx\underbrace{(3/4)^{2/3}\mean{X^2}_c}_{\displaystyle
0.52670}\,\xi^{-2/3},
\label{abe}
\eeq
hence $\mean{r}\sim t/\xi^2$, $\mean{r^2}_c\sim(t/\xi)^{2/3}$.
These results are compatible
with the existence of a smooth crossover between the coarsening
and steady-state regimes, for times of order $t\sim\tau_\eq\sim\xi^4$
[see~(\ref{tauasym})].

\begin{figure}[ht]
\begin{center}
\includegraphics[angle=90,width=.6\linewidth]{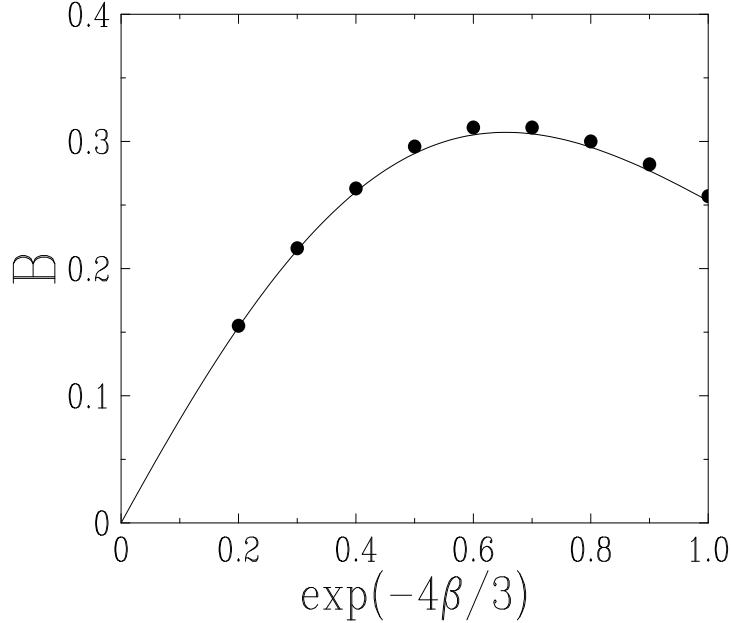}
\caption{\small
Plot of the amplitude $B$ [see~(\ref{lawbe})]
of the spin displacement in the steady-state regime of asymmetric dynamics,
against $\e^{-4\beta/3}$.
Symbols: data.
Full line: prediction~(\ref{kpzb}) of KPZ theory.}
\label{figb}
\end{center}
\end{figure}

Figure~\ref{figb} shows a plot of the amplitude $B$
obtained by numerical simulations, against $\e^{-4\beta/3}$.
Data for lower temperatures are again very costly.
The non-monotonic dependence of $B$ on temperature
is accurately described by the analytical prediction~(\ref{kpzb})
from the KPZ theory.
The data for the drift velocity $V$ are even more accurate,
and indistinguishable from the prediction~(\ref{kpzv}).
At variance with the case of symmetric dynamics,
the values of $V$ and $B$ do not depend on the initial condition.
The measured values indeed coincide with the theoretical predictions,
up to statistical errors, both for an equilibrated
and a disordered initial condition.

\subsection{Displacement variance in the coarsening regime}

We first summarize the known features of the
low-temperature domain-growth process in the presence of a bias~\cite{cb,skr}.
As in Section 3.2, the relevant time scale is $\eps t\sim t/\xi^2$,
since evaporation moves take place with the small probability
$\eps$~(cf.~(\ref{omeg})).

A $(+)$ spin evaporating from the right boundary of a $(+)$ domain
propagates ballistically through the right neighboring~$(-)$ domain,
and condensates to the left boundary of the next $(+)$ domain to the right.
The converse situation of a $(+)$ spin evaporating from the left boundary
of a $(+)$ domain is not permitted.
Integrating over the fast processes (condensation and ballistic propagation)
yields the following description of the dynamics in terms of
domain shifting and annihilation~\cite{cb,skr}.
Each $(+)$ (resp.~$(-)$) domain performs a biased walk
to the right (resp.~to the left),
with a hopping rate $\eps$, irrespective of its length.
When a domain length shrinks down to zero,
the two neighboring domains coalesce.
As a consequence, the whole pattern of~$(+)$ (resp.~$(-)$) domains is advected
with a mean velocity $V$ (resp.~$-V$).
As long as the effective description in terms of domain dynamics holds,
the mean velocity is given by $V\approx\eps$,
just as in the steady-state regime~[see~(\ref{abe})].
The fluctuations around the mean ballistic motion of a given domain
are diffusive, with a diffusion constant $D\approx\eps$.

In the self-similar coarsening regime ($t/\xi^2\gg1$),
the distribution of domain lengths is still of the form~(\ref{dom}).
The behavior of the mean domain length $L$ is now driven by the diffusive
fluctuations.
Since a domain of length $L$ evolves on a time scale
$\eps t\sim L^2$, the following power-law growth holds~\cite{cb,skr}
\beq
L\approx b_L(t/\xi^2)^{1/2},
\label{blpower}
\eeq
where the amplitude $b_L$ is some non-universal number of order unity.

In the coarsening regime, a tagged particle therefore exhibits two
kinds of motion:

\begin{enumerate}

\item[(a)] Most of the time, the tagged particle remains inside a $(+)$ domain.
This domain, and the tagged particle contained in it,
perform a biased random walk, characterized by a mean velocity
and a diffusion constant $V\approx D\approx\eps\approx1/(4\xi^2)$.

\item[(b)] From time to time,
the tagged particle evaporates from the right end
of the $(+)$ domain to which it belongs,
and condensates to the left end of the neighboring $(+)$ domain to the right.
The displacement of the particle in such a move is the length~$\ell$
of the~$(-)$ domain which is crossed.
It is therefore of order $L$.

\end{enumerate}

As in the symmetric case,
moves of either types bring similar contributions to $\mean{r^2}$.
We thus expect, for a given particle and a given initial configuration,
a dispersion of the form
\beq
\qquad\mean{r^2}_c\approx b_r^2\,t/\xi^2,
\label{brpower}
\eeq
for $t\gg\xi^2$, where the amplitude $b_r$ is of order unity.
As a consequence, the dimensionless ratio $\mean{r^2}_c/L^2$
has a non-trivial universal limit value
\beq
R=\lim_{t/\xi^2\to\infty}\frac{\mean{r^2}_c}{L^2}=(b_r/b_L)^2
\label{rr}
\eeq
in the late stages of the coarsening regime.
As the displacement is biased,
all the higher central moments~$K_n$~(\ref{redcm})
are expected to be non-trivial for finite times.
If, however, the dispersion is asymptotically Gaussian,
as it is in the symmetric case,
the cumulants associated with the moments~$K_n$
should fall off to zero in the $t/\xi^2\to\infty$ limit.

\begin{figure}[ht]
\begin{center}
\includegraphics[angle=90,width=.6\linewidth]{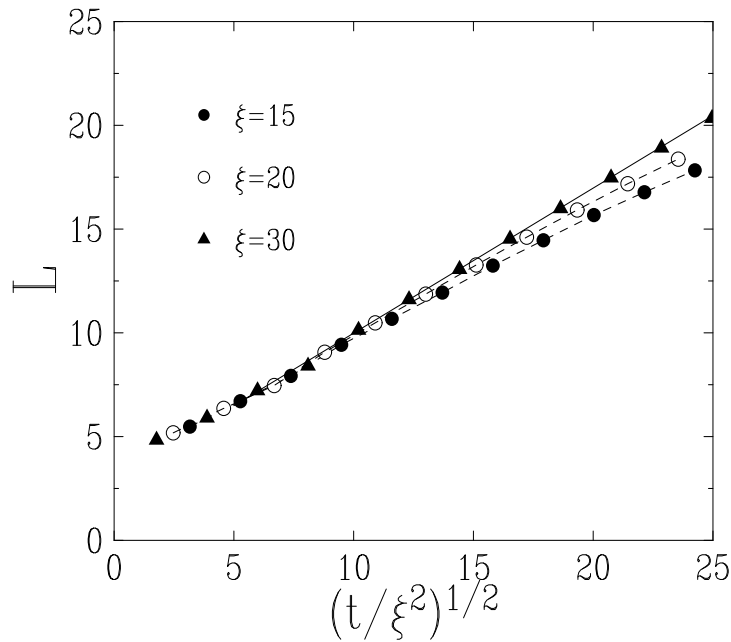}
\caption{\small
Plot of the mean domain length $L$
in the coarsening regime with asymmetric dynamics, against $(t/\xi^2)^{1/2}$.
Symbols: data for several values of $\xi$.
Dashed lines: guides to the eye, emphasizing the corrections to scaling.
Full straight line: least-square fit of the data for $\xi=30$
(first point excluded), confirming~(\ref{blpower}).
The slope yields $b_L\approx0.70$.}
\label{figb1}
\end{center}
\end{figure}

\begin{figure}[ht]
\begin{center}
\includegraphics[angle=90,width=.6\linewidth]{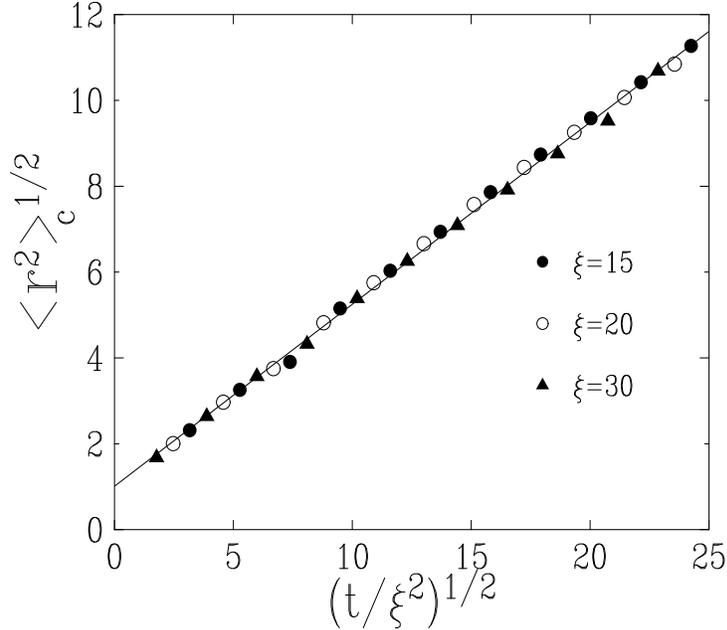}
\caption{\small
Plot of the displacement width $\mean{r^2}_c^{1/2}$
in the coarsening regime with asymmetric dynamics, against $(t/\xi^2)^{1/2}$.
Symbols: data for several values of $\xi$.
Full straight line: least-square fit confirming~(\ref{brpower}).
The slope yields $b_r\approx0.42$.}
\label{figb2}
\end{center}
\end{figure}

\begin{figure}[ht]
\begin{center}
\includegraphics[angle=90,width=.6\linewidth]{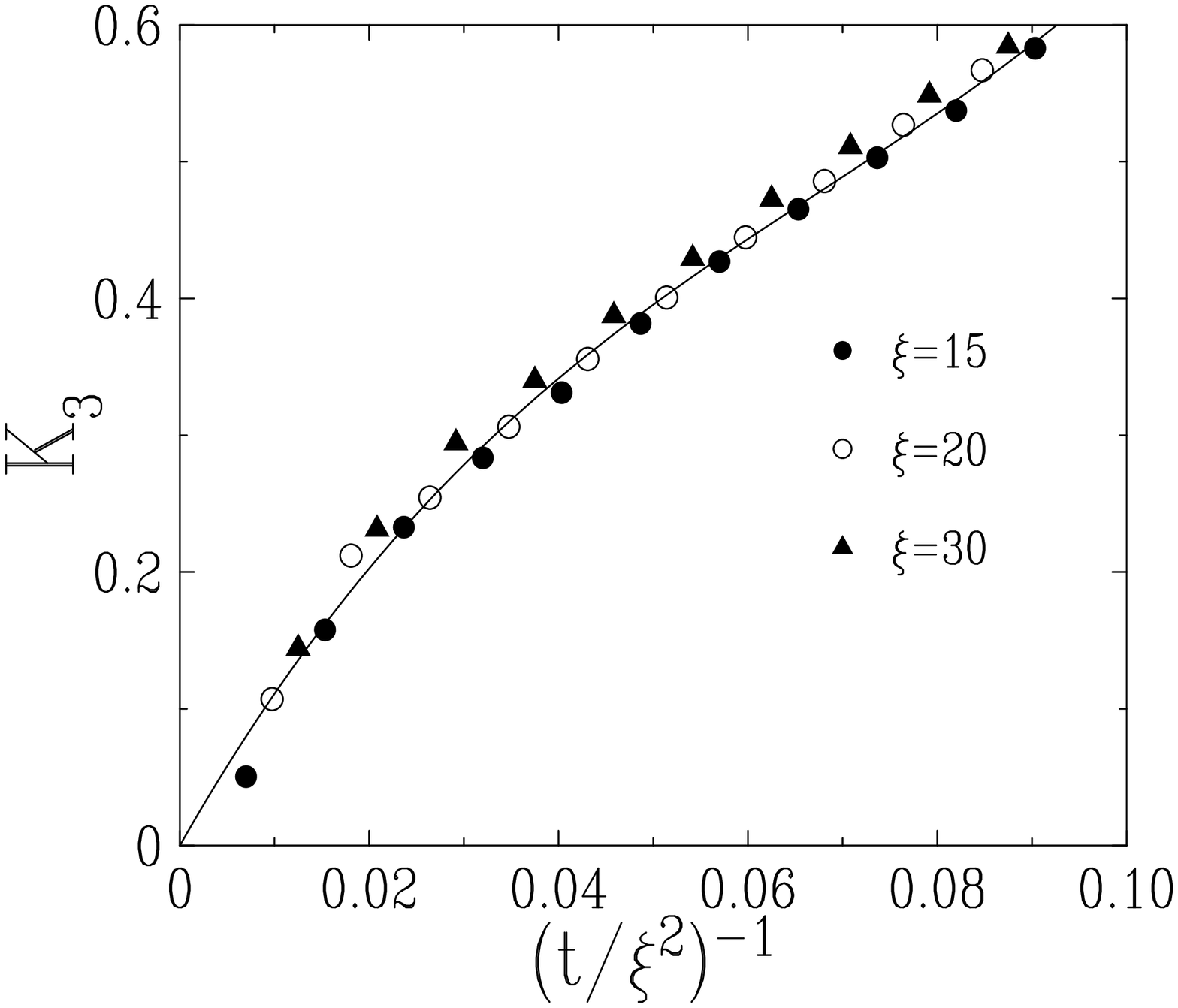}
\caption{\small
Plot of the reduced third central moment (reduced cumulant)
$K_3$ in the coarsening regime with asymmetric dynamics,
against $(t/\xi^2)^{-1}$.
Symbols: data for several values of $\xi$.
Full line: polynomial fit demonstrating the fall-off to zero.}
\label{figb3}
\end{center}
\end{figure}

The following results of numerical simulations
corroborate the above predictions,
and suggest that the distribution of the displacement
is again asymptotically Gaussian.
Figures~\ref{figb1} and~\ref{figb2} show plots
of the mean domain length $L$
and of the displacement width $\mean{r^2}_c^{1/2}$,
against the scaling time variable $(t/\xi^2)^{1/2}$.
The statistics is similar to that of Figures~\ref{figs1} and~\ref{figs2}.
The observed data collapse and linear behavior confirm the
power laws~(\ref{blpower}), (\ref{brpower}).
The data for the mean domain length exhibit appreciable corrections to scaling,
which were hardly visible in the case of symmetric dynamics.
Least-square fits yield
\be
b_L\approx0.70,\qquad b_r\approx0.42,\qquad R\approx0.36.
\ee

Figure~\ref{figb3} shows a plot of the reduced third central moment
$K_3$~[see~(\ref{redcm})] against $(t/\xi^2)^{-1}$.
The observed fall-off to zero strongly suggests
that the dispersion is asymptotically Gaussian
in late stages of the coarsening regime.
We have checked that $K_4-3$ also decays to zero,
albeit with larger statistical errors.

\section{Discussion}

As shown in this work, the motion of a tagged spin
in a ferromagnetic Ising chain evolving under Kawasaki dynamics
is akin to anomalous diffusion.
The variance of the displacement,
$\mean{r^2}_c\equiv\mean{r^2}-\mean{r}^2$, grows algebraically in time,
with an exponent depending on the regime considered.

In the steady state at finite temperature,
the present system is in the universality class
of well-known interacting particle systems,
the SEP in the symmetric case and the ASEP in the asymmetric case,
with subdiffusive behaviors $\mean{r^2}\approx A\,t^{1/2}$
and $\mean{r^2}_c\approx B\,t^{2/3}$, respectively.
Exploiting the relationship with the corresponding continuum theories,
EW in the symmetric case and KPZ in the asymmetric case,
the explicit temperature dependence of the amplitudes
$A$ and $B$ are derived exactly [see~(\ref{aew}) and~(\ref{kpzb})].

Diffusive particle systems with interactions beyond the usual
hard-core repulsion have been the subject of some recent works.
The model for polymer crystallization introduced in~\cite{gw}
is equivalent to asymmetric Kawasaki dynamics
for the antiferromagnetic Ising chain.
In this model, the coupling constant $J$ is negative,
so that $\beta$ is replaced by $-\beta\abs{J}$.
The prefactor $B$ vanishes at a finite critical temperature
such that $\beta\abs{J}=(\ln 3)/2$, as noticed earlier~\cite{kmh}.
Below the critical temperature, the current $J(M)$~(\ref{jofm})
exhibits a minimum in the unmagnetized state $(M=0)$,
and two symmetric maxima for non-zero values of $M$
such that $\exp(-4\beta\abs{J})=1-8/(3-M^2)^2$.
This shape of the density-current characteristic curve
is responsible for the existence of a minimal current phase
in a system with open boundaries~\cite{phase}.
The prototype model studied in~\cite{phase}
is also equivalent to an antiferromagnetic Ising chain,
for a special value of the magnetic field.

In the nonequilibrium low-temperature coarsening regime,
domain growth takes place at the slow time scale $t/\xi^2$.
The effective description of the dynamics on this scale is given
in terms of domain diffusion and annihilation~\cite{cks}
in the symmetric case,
and of domain shifting and annihilation~\cite{cb,skr} in the asymmetric case.
The motion of a tagged particle is very similar in both cases,
the variance of the displacement scaling as the mean square domain length.
Numerical simulations yield the values $Q\approx0.28$, $R\approx0.36$
for the corresponding universal ratios [see~(\ref{qq}) and (\ref{rr})].
The moves of a tagged particle can be of two types, denoted (a) and (b).
The sum of moves (a), where the particle follows a domain,
is clearly asymptotically Gaussian,
whereas moves of type~(b), where the particle jumps
from a domain to another, have a more complex distribution a priori.
Numerical results point however toward a Gaussian dispersion in both cases.

The crossover time between the coarsening regime
and the equilibrium (or steady-state) regime
scales as the equilibration time, which
becomes very large at low temperature $(\xi\gg1)$.
It grows indeed as $\xi^5$ for symmetric dynamics [see~(\ref{tausym})],
and as $\xi^4$ for asymmetric dynamics [see~(\ref{tauasym})].

Table~\ref{res} summarizes
the low-temperature scaling behavior of the mean domain length
and of the variance of the displacement of a tagged spin,
in the steady state and in the nonequilibrium coarsening regime.
For both types of dynamics, the dispersion of a tagged particle
(measured by $\mean{r^2}_c$)
grows faster in the coarsening regime than at equilibrium.
This behavior finds an explanation
once interpreted in terms of a cage effect.
In the coarsening regime, the effective cage felt by a spin
consists of a single domain.
Its size is therefore a typical domain length~$L$, which grows as a power law,
until it saturates to the equilibrium correlation length~$\xi$.
In the steady state, the situation is similar to that of the SEP or ASEP.
The cage effect is a truly many-body effect,
originating in the hard-core exclusion between particles
in a high-density regime, and specific to the one-dimensional geometry.
Remarkably enough, both the equilibrium behavior of $\mean{r^2}$
and the steady-state behavior of $\mean{r^2}_c$ involve the ratio $t/\xi$.

\begin{table}[ht]
\begin{center}
\begin{tabular}{|l|c|c|c|}
\hline
{\bf Symmetric}&$t$&$L$&$\mean{r^2}$\\
\hline
coarsening (Sec.~3.2)&$\xi^2\ll
t\ll\xi^5$&$(t/\xi^2)^{1/3}$&$(t/\xi^2)^{2/3}$\\
equilibrium (Sec.~3.1)&$t\gg\xi^5$&$\xi$&$(t/\xi)^{1/2}$\\
\hline
\hline
{\bf Asymmetric}&$t$&$L$&$\mean{r^2}_c$\\
\hline
coarsening (Sec.~4.2)&$\xi^2\ll
t\ll\xi^4$&$(t/\xi^2)^{1/2}$&$t/\xi^2$\\
steady state (Sec.~4.1)&$t\gg\xi^4$&$\xi$&$(t/\xi)^{2/3}$\\
\hline
\end{tabular}
\caption{\small Dynamical regimes for low-temperature symmetric
and asymmetric Kawasaki dynamics, in the steady state, and
in the nonequilibrium coarsening regime,
with scaling behavior (powers of $t$ and $\xi$) of the typical domain length,
and of the displacement variance of a tagged spin.}
\label{res}
\end{center}
\end{table}

The present work might be extended in several directions.
Let us first mention the crossover between symmetric
and asymmetric dynamics, in the presence of a weak bias.
In the simple case of the mean domain length~\cite{cb},
the essential ingredient of this crossover behavior
is the gambler's ruin problem in the presence of a weak bias.
Another facet of the present problem concerns
the two-time correlation of the displacement,
$C(t+\tau,t)=\mean{(r(t+\tau)-r(t))^2}$.
In the low-temperature coarsening regime,
this correlation function is expected to exhibit
a non-trivial dependence in the two-time plane.

\subsubsection*{Acknowledgements}

It is a pleasure to thank Jean-Philippe Bouchaud
for inspiring discussions at an early stage of this work,
and Kirone Mallick for illuminating conversations.

\end{document}